\documentclass[preprint]{emulateapj}
\newcommand\etal{{\it et al.~}}
\newcommand\msol{\hbox{${\rm M_{\odot}}$}}
\newcommand\ed{\tilde{\delta}}
\newcommand\mach{\mathcal{M}}
\newcommand\cs{c_{\rm s 1}}

\begin{document}

\title{Analytical Models for the Energetics of Cosmic Accretion Shocks,
their Cosmological Evolution, and the Effect of Environment}

\author{Vasiliki Pavlidou\altaffilmark{1,}\altaffilmark{2} and Brian
  D. Fields\altaffilmark{2}}

\altaffiltext{1}{
Kavli Institute for Cosmological Physics, The University of Chicago, 
Chicago, IL 60637}

\altaffiltext{2}{Center for Theoretical Astrophysics, 
Department of Astronomy, 
University of Illinois, Urbana, IL 61801}

\begin{abstract}
We present an analytical description of the energetics of the
population of cosmic {\em accretion} shocks, 
for a concordance cosmology ($\Omega_{\rm m}+\Omega_\Lambda =1$). We calculate
how 
the shock-processed accretion power and mass current are
distributed among different shock Mach numbers, and how they evolve with cosmic
time. We calculate the cumulative energy input of cosmic accretion shocks of 
any Mach number to the intergalactic medium as a function of redshift, and we 
compare it with the energy output of supernova explosions as well as 
with the energy input required to reionize the universe.
In addition, we investigate and quantify the effect of environmental factors, 
such as local clustering properties and filament preheating 
on the statistical properties of these shocks.
We find that the energy processed by accretion shocks is higher than
the supernova energy output for all $z<3$  and that it becomes more 
than an order of magnitude higher in the local universe.
The energy processed by accretion shocks alone becomes comparable to
the energy required to reionize the universe by $z \sim 3.5$. 
Finally, we establish both qualitative and quantitatively 
that both local clustering as well as filament compression and
preheating are important factors in
determining the statistical properties of the cosmic accretion shock
population.

\end{abstract}

\keywords{shock waves -- large-scale structure of universe --
galaxies: clusters:general -- intergalactic medium -- methods: analytical }

\maketitle

\section{Introduction}

The formation of shocks in the baryonic component of matter in the universe is
an inevitable and integral part of the process of cosmological structure
formation. The processing of gas through cosmic shocks
plays a major role in the distribution of cosmic
baryons and their energetics and thermodynamic properties.
In particular, the initial shock heating of primordial
gas leaves it optically invisible and thus contributes
to the inventory of dark baryons (Hegyi \& Olive 1986; Fukugita \etal
1998; Spergel \etal 2003; Cyburt \etal 2003).
A full observational
accounting of dark baryons in all of their states is now
emerging, with the first evidence
for a large component of diffuse, low-density intergalactic gas
distributed in the filaments and sheets that comprise
the ``cosmic web,'' and detected in the X-ray forest 
via absorption lines from
highly-ionized metals
(Hellsten \etal 1998; Perna \& Loeb 1998; Fang \etal 2002;
Nicastro \etal 2002; Finoguenov \etal 2003; Mathur \etal 2003;
Nicastro \etal 2005).
This warm-hot intergalactic medium 
is thought to arise in structure formation shocks 
(e.g. Hellsten \etal 1998; Cen \& Ostriker 1999; Dav\'{e} \etal 2001;
Nath \& Silk 2001; Furlanetto \& Loeb 2004; Kang \etal 2005).

Moreover, structure formation shocks heat the intergalactic medium and 
are likely to act as acceleration sites for nonthermal particles, the 
``structure formation cosmic rays'' 
(e.g. Miniati \etal 2001a,b; Brunetti \etal 2001; Fujita \& Sarazin
2001; Berrington \& Dermer
2003; Gabici \& Blasi 2003a; Brunetti \etal 2004; Kang \& Jones 2005
and references therein).  
Recent detections of nonthermal emission
from galaxy clusters (e.g. Fusco-Femiano \etal 1998; Govoni \etal
2001; Bagchi \etal 2002; Feretti \etal 2004 and references therein)
is consistent with inverse Compton emission 
by relativistic intracluster electrons.
Such a cosmic ray population presumably also includes
hadrons, and so 
would also have distinct $\gamma$-ray and light-element signatures, 
which are currently subjects of intense investigation
(Loeb \& Waxman 2000; Totani \& Kitayama 2000; Miniati 2002; Scharf \&
Mukherjee 2002; Suzuki \& Inoue 2002; Totani \& Inoue 2002; Berrington
\& Dermer 2003; Gabici \& Blasi 2003b, Keshet \etal 2003; Miniati
2003; Reimer \etal 2003; Gabici \& Blasi 2004; Prodanovi\'{c} \&
Fields 2004; Prodanovi\'{c} \& Fields 2005; Suzuki \& Inoue 2004; Kuo
\etal 2005).

Cosmic shocks occur
during different facets of structure formation and  in a variety of
environments, hence 
there are at least two distinct ways to 
categorize them: according to the physical processes
driving them, and according to the state of the medium in 
which they form.

There are three principal processes associated with cosmic structure
formation which result to the formation of large-scale shocks.
\begin{enumerate}
\item Accretion of intergalactic gas by a collapsed, virialized
      structure. In this case, an {\em accretion shock} is formed at
      the interface between virialized and diffuse gas 
      (e.g. Bertschinger 1985a; Miniati \etal 2000)
      The shock is driven by the gravitational  
      attraction exerted on the diffuse gas by the accretor
      and heats the accreted gas to the virial temperature.
\item Merger of two collapsed structures. In this case, a {\em merger
      shock} is formed at the interface between the gas components of
      the merging objects (e.g. Miniati \etal 2000; Gabici \& Blasi 2003).
      The shock is driven
      by the mutual gravitational attraction between the objects.
\item Accumulation of intergalactic gas onto large-scale cosmic
      filaments and sheets.  Contrary to virialized structures,
      large-scale filaments and sheets, although overdense, are still
      expanding. At the edges between filaments and their
      surroundings, shocks form, driven by the difference in expansion
      velocity between the overdense filament and the underdense
      neighboring voids (e.g. Bertschinger 2005b).
      More
      complicated shock surfaces inside the filaments are additionally 
      produced due to the gravity of the collapsed objects residing in
      the filament interior.
\end{enumerate}

Shocks can also be divided according
to the state of the gas passing through them, into {\em external} and
{\em internal} shocks (e.g. Miniati \etal 2000; Ryu \etal 2003).
External shocks process pristine material, which has never been
shocked before by any of the processes described above. External
shocks are mostly filament shocks, since the process of formation of
individual virialized structures (associated with the other two types of
shocks) occurs principally within filaments, and therefore in most
cases involves gas which has already been processed at least by filament
shocks. Internal shocks process gas which has already been shock-heated in the
past. All merger shocks, as well as many accretion shocks, are internal shocks.

Since external shocks process colder material of 
lower sound speed, their Mach numbers are generally higher
than those of internal shocks.
However, because the gas passing through internal shocks has 
already been compressed, internal shocks
process more mass and kinetic energy than external shocks.

In this work, we present an analytical study of the energetics and 
cosmic evolution of the population of  
of {\em cosmic accretion shocks}, in a concordance $\Omega_{\rm
m}+\Omega_\Lambda=1$ universe. In addition, we assess 
the effect of environmental factors on the properties of these
shocks. The environmental factors we consider are: (a) 
variations in the local matter density and temperature
of the region in which each accretor is embedded imprinted in the
primordial density field and (b) filament preheating and compression.

In order to distinguish between the effects of each environmental
factor, we start from a base model which accounts for no environmental
effects and regards all object as being embedded in an environment
well represented by the background universe. In this case, shock
properties depend solely on the mass of the accretor.
To this model we then add each of the environmental factors mentioned 
above, each version of our model being refined by one effect with 
respect to the previous version. The final (full) model we construct 
in this way accounts for the effects of the mass distribution of
accretors as well as both environmental factors considered here. 

Although each effect is treated under a number of simplifying
assumptions which necessarily introduce some error in our
calculations, we make sure to on one hand quantify this error whenever
possible and on the other hand to consistently err on the side of {\em
underestimating} the effect of environment. In this sense, our
results are an absolute lower limit of the effect of environmental
factors on the properties of the population of cosmic accretion
shocks. In addition, the {\em trends} and {\em qualitative changes}
introduced to the properties of the accretion shock population are
clearly identified. Any future quantitative correction to our model
will only enhance such trends. 

The shock properties as a function of object mass are derived from 
the temperature jump across the shock surface (as is the case in
cosmological numerical simulations), while the underlying mass-function
of the accretors is taken to be the Press-Schechter mass function.
We calculate statistical quantities characteristic of the population of cosmic 
accretion shocks, such as the distribution of number density of 
objects, 
processed kinetic energy and shocked mass with respect to
Mach number, as well as their evolution with cosmic time, and we use
our findings to discuss the relative importance of cosmic accretion
shocks in the energetics of the intergalactic medium. 

Our paper is organized as follows. 
In \S \ref{sshock} we present the
formalism describing  
the properties of a single accretion shock around a cosmic structure. 
In \S \ref{popshocks} we combine our
single-shock model with different cosmological distribution functions
to derive the statistical properties of the population of cosmic
accretion shocks. In \S \ref{results} we present the results of our
model for a concordance $\Omega_{\rm m}+\Omega_\Lambda=1$ universe.
In \S \ref{pastwork} we discuss
the differences and similarities of our approach from other analytic
cosmic shock models based on the Press-Schechter approach.
Finally, we summarize and discuss our findings in \S \ref{disc}.  

\section{Properties of a single shock}\label{sshock}

In this work, 
the properties of the accretion shock around a single virialized host
object are derived from the temperature jump across the shock, in
agreement with the method used to derive shock properties in
cosmological simulations. 
The pre-shock temperature is taken to be an appropriately defined
environmental temperature, while the post-shock temperature is
taken to be the virial temperature of the host structure. However, the
virial temperature is a {\em mean} quantity of the host object and
does not necessarily characterize the temperature right behind the
shock. To assess the error introduced in our calculation of the Mach
number due to this assumption, we compared our temperature-jump
technique with the Bertschinger (1985) similarity 
solution which {\em does} take into
account the radial dependences of the post-shock gas. To ensure an
appropriate comparison, we applied our formalism to an Einstein-de
Sitter universe and in the high-Mach limit, where the Bertschinger
solution is applicable. We found that the deviation from the
Bertschinger solution is at the level of 1$\%$. 

Throughout this paper, we assume an adiabatic equation of state, and
we consider all shocks to be non-radiative.  We also assume that any
individual collapsed object as well as its accretion shock are
spherically symmetric.
We take the accretion shock position
around each structure to coincide with the virial radius of each
structure.

The Mach number of a shock, $\mach$, is defined as the ratio of the
velocity of the accreted material in the shock frame to the adiabatic sound
speed of the accreted material.
The Mach number is related to the temperature jump across the shock through
\begin{equation}\label{jumpspec}
\frac{T_2}{T_1}=\frac{(5\mach^2-1)(\mach^2+3)}{16\mach^2}\,.
\end{equation}
where $T_1$ and $T_2$ are the pre-shock and post-shock temperatures
correspondingly, and we have assumed a ratio of specific heats
$\gamma=5/3$, constant across the shock. 
In the limit $\mach \gg 1$ this equation is further simplified, 
$\mach = \sqrt{(16T_2)/(5T_1)}$.
The pre-shock temperature can be written in terms of the adiabatic
sound speed of the pre-shock material $\cs$ as  
$ kT_1 = \mu m_{\rm p}\cs^2 /\gamma\,,$
where $k$ is the Boltzmann constant, 
$\mu$ is the mean molecular weight of the accreted gas, and 
$m_{\rm p}$ is the proton mass.  If we also take 
$T_2$ to be the virial temperature of the accretor which has a mass $m$, 
$ T_{\rm vir} = \mu m_{\rm p} 
Gm^{2/3}(4\pi f_{\rm c} \rho_{\rm m,0})^{1/3}(1+z) /
(5k3^{1/3})$, 
then the ratio $T_2/T_1$
becomes
\begin{equation}\label{jumpvir}
\frac{T_2(m,z)}{T_1}  =
2.7 \times 10^3\Omega_{\rm m}\left(\frac{f_c}{18\pi^2}
\frac{m^2}{m_8^2}\right)^{1/3}
\!\!\!\!\!\!(1+z) \!\left(\frac{15 {\rm \,\, km \,s^{-1}}}{\cs}\right)^2,
\end{equation}
where $f_c$ is the compression factor for a virialized object (which
may vary with virialization redshift, depending on the cosmological
model), $z$ is the virialization redshift, 
$h$ is the dimensionless Hubble parameter, and $m_8 = 5.96 \times
10^{14} h^{-1} \Omega _{\rm m} \msol$ is the mass included in a sphere
of comoving radius $r_8=8 h^{-1} {\, \rm Mpc}$ assuming the mean matter density
inside the sphere to be equal
to the cosmic mean.

The (baryonic) mass current, defined as the rate at which mass crosses the
surface of a single accretion shock
around a structure of mass $m$ at an epoch $z$, is
\begin{equation}\label{j1eq}
J_1 = \frac{\Omega_{\rm b}}{\Omega_{\rm m}}\frac{dm}{dt} =  
4\pi r_{\rm v}^2(m)\Omega_{\rm b}\rho_{\rm c,0}(1+z)^3 (1+\delta_{\rm s})
\mach \cs
\end{equation}
where $r_{\rm v}$ is the virial radius of the structure, 
\begin{equation} \label{rvir}
r_{\rm v}=
 1.4 h^{-1} {\rm \, Mpc }
\left(\frac{m}{m_8}\right)^{1/3}
\left(\frac{f_{\rm c}}{18\pi^2}\right)^{-1/3}
(1+z)^{-1}\,,
\end{equation}
$\rho_{\rm c,0}$ is the critical density at the present cosmic epoch, 
$\Omega_{\rm b}\rho_{\rm c,0}(1+z)^3$ is the cosmic baryon density 
at the epoch of
interest, and $(1+\delta_{\rm s})$ is the density enhancement 
(with respect to the cosmic mean) just outside the shock.

Finally, the kinetic power crossing
  the accretion shock around a single structure of mass $m$ is
$P_1 = dE/dt = 0.5(dm/dt)v_{1}^2 = 
 0.5J_1v_{1}^2$, or
\begin{equation}\label{p1eq}
P_1 = 2 \pi r_{\rm v}(m)^2 \Omega_{\rm b}\rho_{\rm c,0}(1+z)^3
(1+\delta_{\rm s}) \mach^3 \cs^3 \,.
\end{equation}

\section{Properties of the population of cosmic accretion 
shocks}\label{popshocks}

We will examine three distinct models for the cosmic accretion shock
population. We will start from a ``base'' no-environmental-effects model,
in which the properties of each accretion shock are simply a function
of the accretor mass. We will then gradually build up more complex
models by adding one-by-one
the different environmental factors we wish to examine (local
density and temperature variations originating in the primordial
density field, and filament compression and preheating). In this
way, we will assess the effect of each factor on the properties of the
cosmic shock population. Our third, most complex model, will include
the effect of both environmental factors we are addressing here. 

\subsection{Model 1 (base model): no environmental effects}

If we assume that all collapsed 
objects accrete baryons of uniform density and temperature (the
density and temperature corresponding to the mean, background-universe
values of diffuse baryons), then at a given redshift, all objects 
of a certain mass will have accretion shocks with identical Mach
numbers, and will process the same amount of mass and kinetic power
per unit time. In the relations of \S \ref{sshock}, $\delta_{\rm s}=0$
and $\cs=15 {\rm \, km/s}$ as we will be treating only
post-reionization redshifts. The mass distribution of collapsed objects
will be the one described by the Press-Schechter mass function 
(Press \& Schechter 1974; Lacey \& Cole 1993)
\begin{equation}\label{psmf}
\frac{dn}{dm}(m,z) = \sqrt{\frac{2}{\pi}}\frac{\rho_{m,0}}{m^2}
\left|\frac{d\ln \sigma}{d\ln m }\right|
\exp\left\{-\frac{\left[\tilde{\delta}_c(z)\right]^2}{2\left[\sigma(m)\right]^2
}\right\} \,.
\end{equation}
where $\tilde{\delta}_c(z)$ is the linearly extrapolated overdensity
of an object which collapses at redshift $z$, $\sigma(m) =
\sqrt{S(m)}$ 
is the
square root of the variance of the
linearly extrapolated field smoothed at a mass scale $m$, and $\rho_{m,0}$ is
the cosmic mean matter density at the present time.

\subsection{Model 2: model 1 + effect of primordial density field}

In the second variation of our model, we will relax the assumption that the
temperature of the accreted material and the pre-shock density 
is the same for all structures, and we will consider the effect of
variations in the local environment of accretors of the same mass
caused by the structure of the primordial density field. In this case,
accretors of the same mass will reside in a distribution of environments,
parametrized by the  local overdensity or underdensity (as determined
by the evolution of primordial density fluctuations), $
\delta_\ell = (\rho_{\rm local} - \rho_{\rm m})/\rho_{\rm m}$.
In addition to affecting the pre-shock density enhancement,
$\delta_{\rm s}+1 = \delta_{\ell}+1$, such density variations will also cause
temperature variations of the pre-shock gas, 
due to adiabatic heating or cooling. Since  $c_{\rm s}^2\propto \rho_{\rm
  local}^{\gamma-1}$, the pre-shock sound speed will in this case be
  a function of $\delta_\ell$, 
\begin{equation}\label{thecs}
c_{\rm s}= c_{\rm s,avg} \left(\delta_\ell+1\right)^{1/3}\,, 
\end{equation}
for our $\gamma=5/3$ gas. In eq. (\ref{thecs}),
$c_{\rm s,avg}$ is the ``cosmic average'' sound speed 
(the sound speed of the intergalactic
medium at a density equal to the cosmic mean at
 the epoch of interest), which we will again take to be equal to $15
 {\rm \, km/s}$. 

In order to make further progress and be able to calculate measures of
the statistical properties of the population shocks in this
approximation, we need an analytical model for the {\em environment}
of collapsed structures. For this purpose, we will use the double
distribution (DD) of collapsed structures with respect to mass and local
overdensity (Pavlidou \& Fields 2005, hereafter PF05).
In PF05,  we defined the ``local
environment'' of a collapsed structure through the {\em clustering
  scale parameter}, $\beta$. The clustering scale parameter is a free
parameter in the double distribution model, and is 
defined so that  the ``environment'' of an object of mass $m$ be a
surrounding region in space which encompasses mass $\beta m$. Clearly,
as $\beta \rightarrow 1$ the environment of an object is restricted
to be closer to the object itself, and for $\beta=1$ it encompasses
solely the collapsed object itself. Conversely, as $\beta \rightarrow
\infty$, the environment of the object extends further and further
away from the object, to eventually encompass the whole universe.

With this definition of the ``local environment'', and using the same
random walk formalism which yields the Press-Schechter mass function
for collapsed objects, we found the DD to be 
\begin{eqnarray} \label{dd}
\frac{dn}{dmd\ed_\ell}
&=&
\frac{\rho_{m,0}}{m} \,\,
\frac{\ed_c(a)-\ed_\ell}{2\pi } \left|\frac{dS}{dm}\right|_m
\!\! \exp\left[-\frac{\left(\ed_c(a)-\ed_\ell\right)^2}
{2\left[S(m)-S(\beta m)\right]}\right]
\nonumber \\
&& \times \frac{
\exp \left[-\frac{\ed_\ell^2}{2S(\beta m)}\right]
- \exp\left[-\frac{\left(\ed_\ell - 2 \ed_c(a)\right)^2}{2S(\beta
    m)}\right]
}
{[S(\beta m)]^{1/2}
\left[S(m)-S(\beta m)\right]^{3/2}}
\end{eqnarray}

In Eq. (\ref{dd}), $S(m)$ is the variance of the linearly
extrapolated overdensity field
and  $\ed_\ell$ is the local linearly extrapolated
overdensity (or underdensity), which is related to the true 
(calculated from the
spherical evolution model) overdensity  $\hat{\delta}_{\ell}$ 
of the environment sphere {\em
  including the local object}
through the exact relations
given in PF05 or through the useful approximation (accurate at
a better than $2\%$ level throughout its domain for all cosmologies of
interest) 
\begin{equation}\label{apconv}
\ed_\ell \approx
\frac{D(a_0)}{D(a)}\ed_c\left[1-(1+\hat{\delta}_\ell)^{-1/\ed_c}\right]\,, 
\end{equation}
where $D(a)$ is the linear growth factor in the cosmology of interest
and $\ed_c$ is the linear overdensity threshold for collapse in the
same cosmology. The quantity $\hat{\delta}_\ell$ is in turn related to the
overdensity of interest, $\delta_\ell$ 
(the overdensity of the environment sphere
{\em excluding the central object}), through 
\begin{equation}
\delta_\ell = \frac{(\beta-1)(1+\hat{\delta}_\ell)(1+\delta_c)}
{\beta(1+\delta_c)-(1+\hat{\delta}_\ell)}-1\,.
\end{equation}

Because in the problem of cosmic accretion shocks we are interested in 
the properties (density and sound speed) of the material right outside
the shock surface, we will adopt a small value for the clustering
scale parameter, $\beta=1.1$. Our results, however, are not sensitive
to the exact value of $\beta$ since, as we found in PF05, the
properties of the double distribution (when
calculated as a function of $\delta_\ell$ rather than $\hat{\delta}_\ell$) 
depend only mildly on $\beta$ for small values of $\beta$.

\subsection{Model 3 (full model): model 2 + effect of filaments}

In this third variation of our model, we will expand the
double-distribution--based model 2 to include the effect of filament
preheating and compression. Most virialized structures reside inside
filaments, and hence they are accreting gas which has already
compressed and preheated in filament shocks. A complete description of
the effect of filaments on the properties of cosmic accretion shocks
would require an analytic model for the filamentary structure of the
universe accounting for a distribution of 
different filament temperatures and densities, as well as
their evolution with redshift, a problem which we will 
address in detail in an upcoming publication. 
Here, we adopt a simplified prescription which can
provide a first assessment of the degree to which filaments alter the
energetics and evolution of cosmic accretion shocks. 

We will assume
that all structures with environmental overdensity $\delta_\ell > 0$
reside inside filaments\footnote{This is a conservative
assumption. In reality, even structures predicted by the double
distribution model to reside inside underdensities may be found inside
filaments, as neighboring voids may sufficiently compress their
surrounding regions.}. These structures, instead of accreting gas
of mean sound speed $c_{\rm s,avg}$, will be accreting
filament-preheated gas of mean sound speed $c_{\rm s, fil} \approx 45
{\, \rm km /s}$,
corresponding to a (conservative) mean filament
temperature $T_{\rm fil} \approx 10^5 {\, \rm K}$ (e.g. Cen \& Ostriker
1999, Dav\'{e} \etal 2001; Fang \etal 2002).
In addition, $1+\delta_\ell$ will now represent the
compression factor with respect to a mean filament density $\rho_{\rm
fil} = \rho_{\rm b} (1+\delta_{\rm fil})$ where (again,
conservatively)\footnote{Note that our prescription does not account for
  filament evolution with cosmic time. Of course, filaments {\em do}
  evolve with redshift, and a more detailed calculation, based on a
  model accounting for filament evolution as well as different
  filament overdensities and temperatures is desirable and currently
  pursued. However, the error introduced by not accounting for
  filament evolution is expected to be relatively small, as filaments primarily
  evolve by shock-heating and compressing increasing {\em quantities} of
  diffuse gas, rather than heating it and compressing it to
  significantly different densities and temperatures (see e.g. distributions in
  Dav\'{e} \etal 2001, which change in amplitude rather than shift in
  location with changing redshift). This however is not expected to
  significantly affect the properties of gas accreted by virialized
  structures: filaments may include smaller quantities of gas at high
  redshifts, but structures are also smaller in size and most of them 
  are still expected to be found inside filaments.} $1+\delta_{\rm fil}
\approx 10$. Hence, in model 3 the
pre-shock sound speed will be 
\begin{equation}\label{mod3cs}
c_{\rm s}= \left\{
\begin{array}{lr} 
c_{\rm s,avg} \left(\delta_\ell+1\right)^{1./3}\,, & \delta_{\ell}\leq
0\\
c_{\rm s,fil} \left(\delta_\ell+1\right)^{1./3}\,, 
& \delta_{\ell}>0  
\end{array}
\right.\,,
\end{equation}
while the pre-shock density enhancement will be 
\begin{equation}\label{mod3ds}
1+\delta_{\rm s} = \left\{
\begin{array}{lr}
1+\delta_\ell \,, & \delta_\ell \leq 0 \\
(1+\delta_{\rm fil})(1+\delta_\ell)\,, & \delta_\ell > 0
\end{array}
\right.\,.
\end{equation}
The distribution of number density of accretors with respect to mass
and local overdensity $\delta_\ell$ will again be assumed to be well
described by the double distribution of collapsed structures.

\subsection{Quantitative measures of the properties of cosmic
accretion shocks}

The specific properties of the population of cosmic accretion shocks
which we wish to investigate can be classified into two
categories. First of all, we are interested to see how the
shock-processed mass and energy is distributed among shocks of
different Mach numbers, and how environmental effects alter this
distribution. Such Mach number distributions are of particular importance
to assess the role of cosmic accretion shocks as potential sites of 
particle acceleration, since the highest-energy particles can be
accelerated only in the strongest ($\mach \gg 1$) shocks. 
The quantities we will calculate to address this question are: 
\begin{itemize}
\item The ``mass current distribution'' with respect to Mach number,
      $dJ/d\ln\mach$,  
defined as the  comoving mass current density crossing shock 
surfaces of logarithmic Mach number between $\ln \mach$ and $\ln \mach+d\ln
  \mach$,  
with units of $\msol {\rm \, yr^{-1}\, Mpc^{-3}}$. In the case of
model 1, it is given by 
\begin{equation}\label{psfirst}
\frac{dJ}{d\ln\mach} = \mach 
\frac{dn}{dm}\left. \frac{\partial m}{\partial \mach}\right|_z
J_1\,,
\end{equation}
where $m=m(\mach,z)$ is given by eqs.~(\ref{jumpspec}) and
(\ref{jumpvir}) with $\cs = 15 {\rm \, km/s}$; $dn/dm$ is the
Press-Schecter mass function, given by eq.~(\ref{psmf});
and $J_1$ is given by eq.~(\ref{j1eq}) with $\delta_{\rm
s} =0$. In the case of models 2 and 3, it is given by
\begin{equation}
\frac{dJ}{d\ln\mach}=\mach \int_{\delta_\ell=-1}^{\delta_c} d\delta_\ell
J_1 \frac{dn}{dm d\delta_\ell}\left.\frac{\partial m}{\partial
  \mach}\right|_{\delta_\ell,z}\,,
\end{equation}
where $dn/(dm \,d\delta_\ell)$ is the double distribution of
cosmic structures, given by eq.~(\ref{dd}); 
$m(\mach, \delta, z)$ is again given by eqs.~(\ref{jumpspec})
and (\ref{jumpvir}), with $\cs$ given by eqs.~(\ref{thecs}) and
(\ref{mod3cs}) for models 2 and 3 respectively; and
$\delta_{\rm s} = \delta_\ell$ for model 2, while
$\delta_{\rm s}$ given by eq.~(\ref{mod3ds}) for model 3.

\item The ``kinetic power distribution'' with respect to Mach
      number, $dP/d\ln \mach$,  
defined as the comoving kinetic power density crossing 
shock surfaces of
logarithmic Mach number between $\ln \mach$ and $\ln\mach+d\ln\mach$, 
 with units ${\rm erg \, s^{-1} \,}$ ${\rm Mpc^{-3}}$.
It is given by 
\begin{equation}\label{pslast}
\frac{dP}{d\ln \mach} = 
\mach 
\frac{dn}{dm}\left. \frac{\partial m}{\partial \mach}\right|_z
P_1
\end{equation}
for model 1 and by 
\begin{equation}
\frac{dP}{d\ln \mach} = \mach \int_{\delta_\ell=-1}^{\delta_c} d\delta_\ell
P_1\frac{dn}{dm d\delta_\ell}\left.\frac{\partial m}{\partial
  \mach}\right|_{\delta_\ell,z}\,,
\end{equation}
for models 2 and 3. In both cases, $P_1(\mach,z)$ is given by Eq.~(\ref{p1eq}).
\end{itemize}

Second, we would like to follow the cosmic evolution of the energetics
of cosmic accretion shocks, tracing their mass and energy processing and their 
impact on the intergalactic 
medium. In particular, we wish to calculate the energy amount
processed by all cosmic accretion shocks per unit time and by a
certain redshift, as well as the degree to which cosmic baryons have
been affected by cosmic accretion shocks by some specific
cosmic epoch. The quantities we will use to quantify these questions
are:

\begin{itemize}
\item The ``integrated mass current density'', 
$J$, which is the comoving mass current
      density crossing shock surfaces of {\em any} Mach number at a given
      cosmic epoch, with units of $\msol {\rm \, yr^{-1}\, Mpc^{-3}}$.

\item The ``integrated kinetic power density'', 
$P$, which is the comoving kinetic
      power density crossing shock surfaces of any Mach number at a
      given cosmic epoch, with units of ${\rm erg \, s^{-1} \,}$ 
${\rm Mpc^{-3}}$.

\item The ``cumulative processed kinetic energy density'', 
$\int _{t_{\rm
      i}}^{t} P dt$, which is the total kinetic energy density 
      processed by shocks
      of any Mach number since some initial cosmic epoch $t_{\rm i}$,
      with units of ${\rm eV}$ per baryon in the universe. 

\end{itemize}

The second set of quantities will be calculated as a function of
redshift, and for each of the models 1, 2 and 3, in order to assess 
the effect of each environmental factor on their behavior.

\section{Results}\label{results}

Throughout this section we will 
assume a flat, vacuum+matter universe with 
 {\it Wilkinson Microwave Anisotropy Probe} 
(WMAP) parameters $\sigma_8=0.84$,
$h=0.71$, $\Omega_{\rm m}=0.27$ and $\Omega_{\rm b}=0.04$ 
(Spergel \etal 2003).
We focus on post-reionization redshifts, hence we adopt a
cosmic average sound speed of
$15 {\rm \, km/s}$, corresponding to a temperature of $\sim 10^4$K,
for a fully ionized plasma with $\mu=0.59$ ($25\%$ He by mass). 

\subsection{Mach number distributions and the effect of the environment}
\label{analarg}

\begin{figure*}
\plottwo{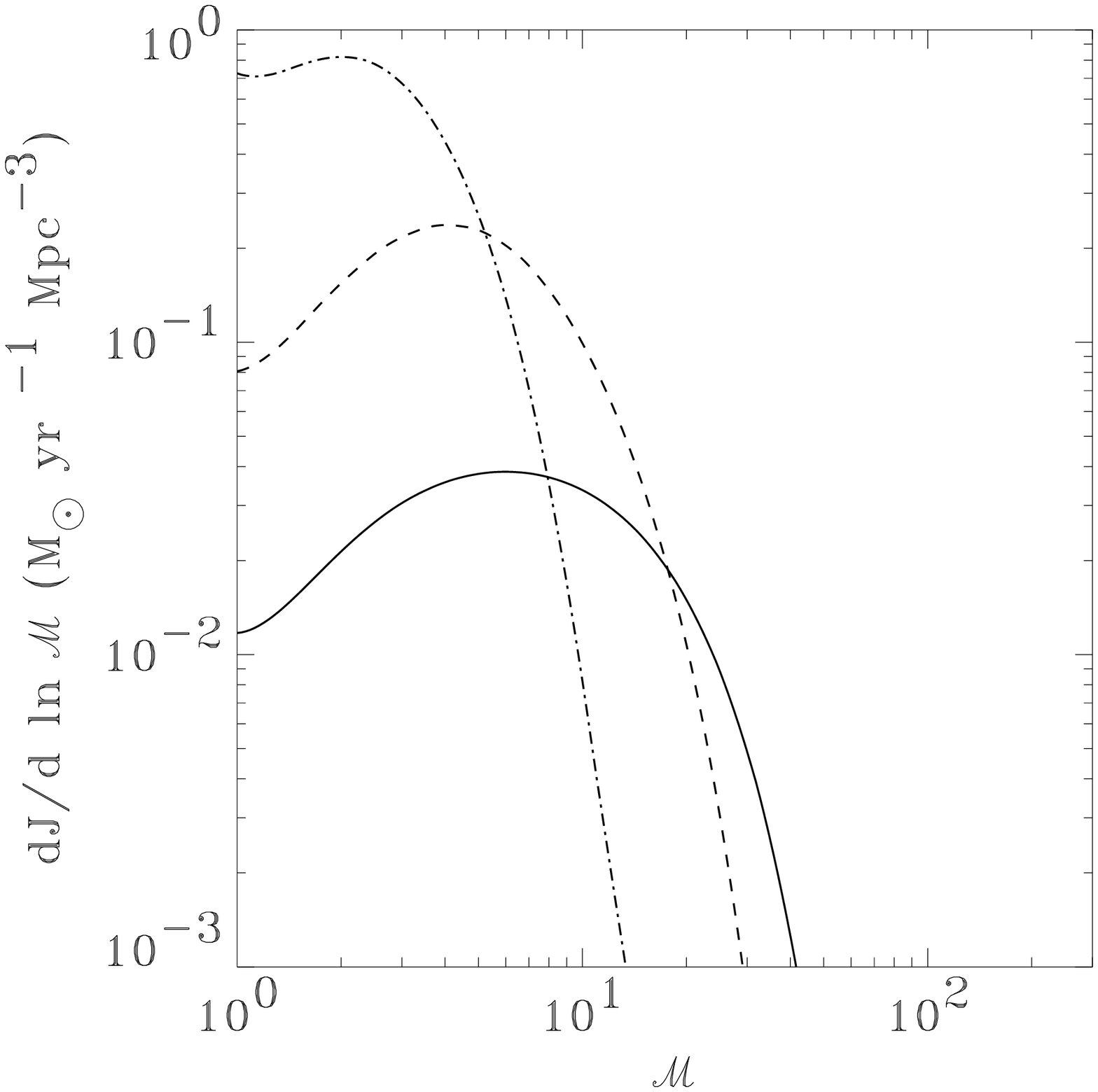}{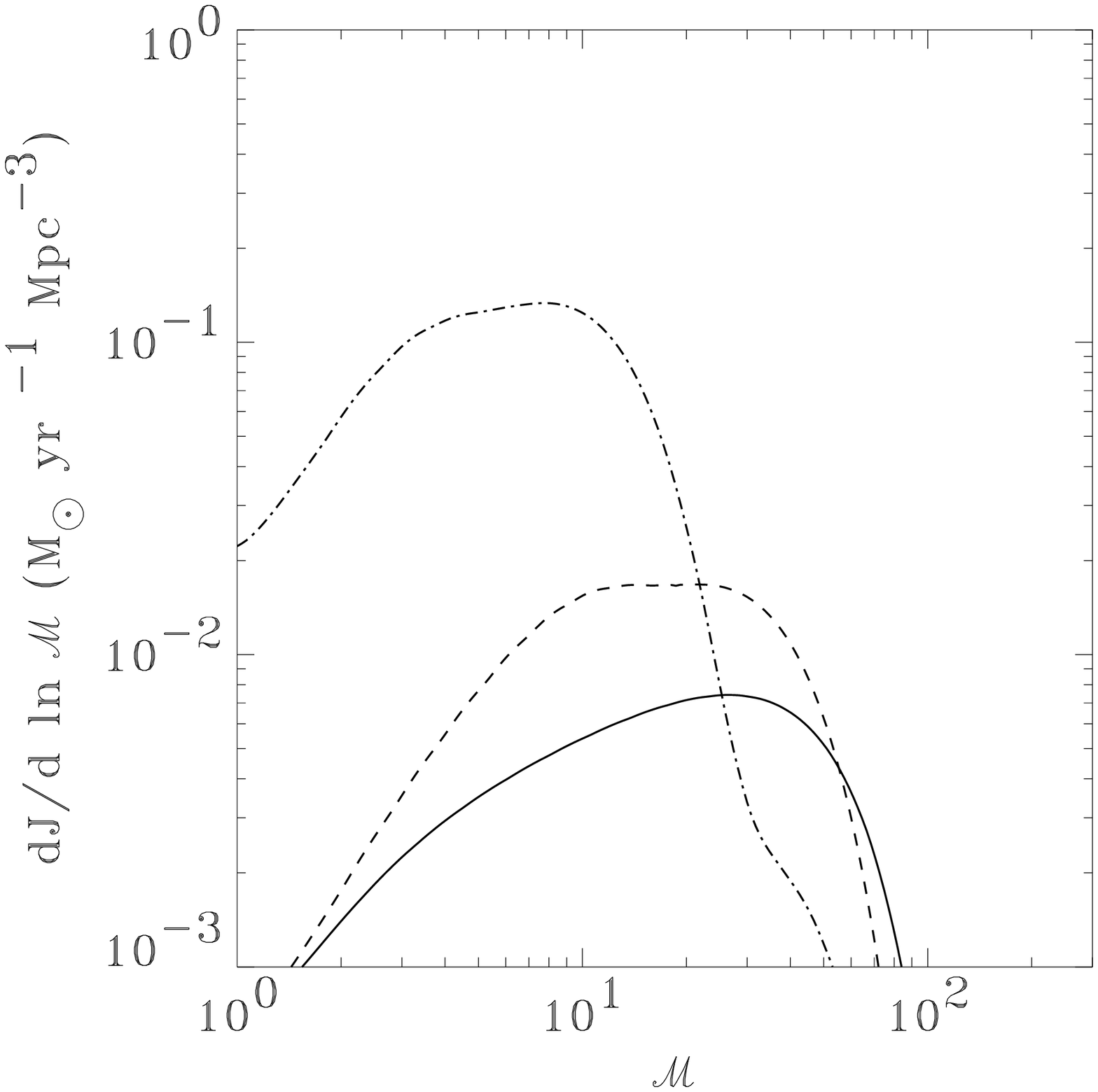}
\caption{\label{fig:mcdists} 
Mass current distribution (spatial density of
mass accretion rate 
per logarithmic Mach number interval, $dJ/d\ln\mach$)for $z=3$
(left panel) and $z=0$ (right panel). 
The units of the vertical axes are ${\rm M_\odot \,\,
 yr ^{-1} \, comoving \, Mpc^{-3}}$.
Solid line: model 1 (no environmental effects); dashed line: model 2
(effect of primordial density fluctuations); dot-dashed line: model 3
(effects of primordial density fluctuations and filaments).}
\end{figure*}

\begin{figure*}
\plottwo{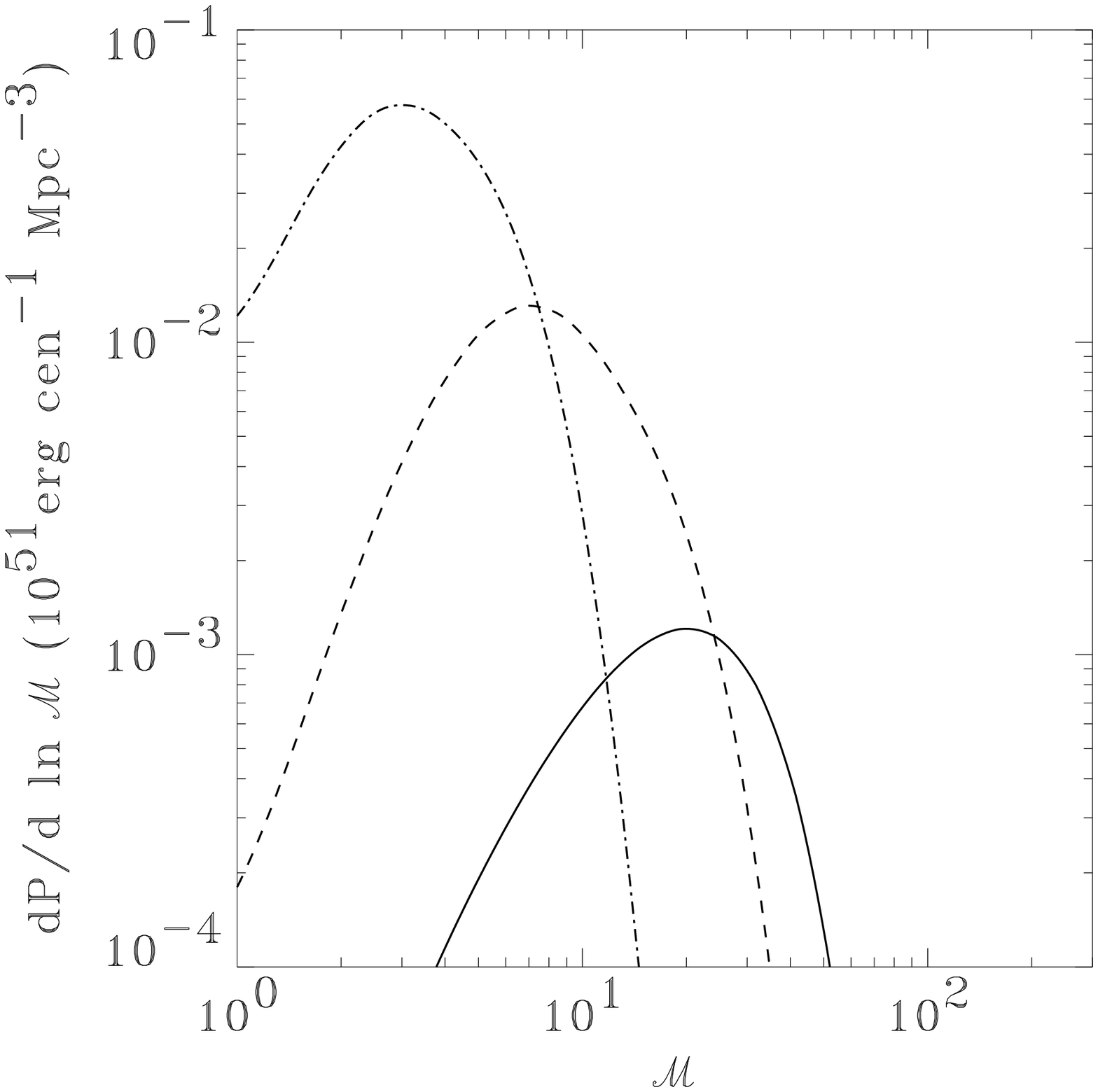}{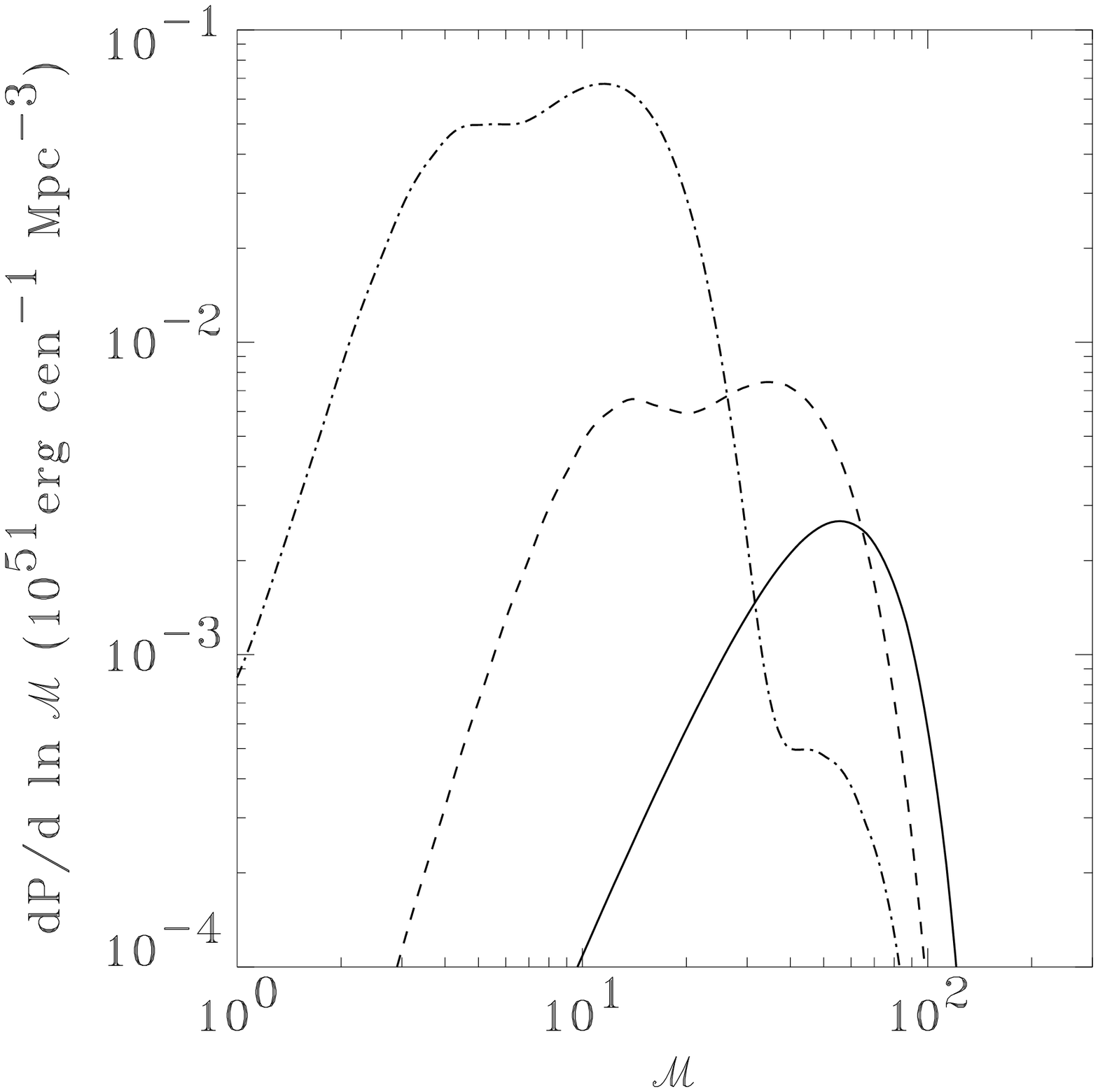}
\caption{\label{fig:kindists} 
Kinetic power distribution (spatial density of
 kinetic power processed by accretion shocks
per logarithmic Mach number interval, $dP/d\ln\mach$)for $z=3$
(left panel) and $z=0$ (right panel). 
The units of the vertical axes are ${\rm g \,\,
 s ^{-1} \, comoving \, Mpc^{-3}}$.
Solid line: model 1 (no environmental effects); dashed line: model 2
(effect of primordial density fluctuations); dot-dashed line: model 3
(effects of primordial density fluctuations and filaments).}
\end{figure*}

The distributions with respect to Mach number of mass current and
kinetic power are plotted in figs. \ref{fig:mcdists} and \ref{fig:kindists}
respectively. In both figures, the solid line corresponds to model 1
(Press-Schechter--based, no environmental effects); the dashed line
corresponds to model 2 (double-distribution--based, includes the
effect of the primordial density field but not the effect of
filaments); and the dot-dashed line corresponds to model 3 (includes
effects of both primordial density field as well as effects of
filaments). The left panel shows the results for $z=3$, while the
right panel corresponds to the present-day
universe.

The results of the Press-Schecter--based, no-environmental--effects
model 
can be interpreted in a straight-forward way. In this
case, accretor mass and Mach number have a one-to-one correspondence,
with larger accretor masses resulting to higher--Mach-number accretion
shocks. As cosmic time progresses, the mass of the largest collapsed
objects increases, as more and more massive structures have had enough
time to collapse. Hence, the strongest accretion shocks become
stronger with decreasing redshift, and the distributions of mass
current and kinetic energy move to higher Mach numbers with increasing time.
The kinetic power distribution is more strongly suppressed at low Mach
numbers (and therefore low accretor masses) than the mass 
current distribution. This result can be immediately verified by
simple analytic arguments.
For masses high enough that the primordial density fluctuation 
power spectrum can be regarded as a power law but low enough that the 
exponential mass cutoff is not affecting the results,  
$dn/dm \propto m^{-2}$. In addition, $\mach \propto m^{1/3}$ (in the
high-$\mach$ limit), 
while $r_{\rm v} \propto m^{1/3}$.
Hence, $J_1 \propto m$ and $P_1 \propto m^{5/3}$. Equations
(\ref{psfirst}) and (\ref{pslast}) then give for the low-mass dependence of
the mass current and kinetic power distributions, 
\begin{eqnarray}
\frac{dJ}{d\ln \mach} &\propto &m^0 \propto \mach^0 \nonumber \\
\frac{dP}{d\ln \mach} &\propto& m^{2/3} \propto \mach^2 \,
\end{eqnarray}
accounting for the difference in the low-mass behavior of different
distributions. 
Note that the low-mass suppression is somewhat stronger
than what predicted by the simple arguments above, due to the deviation
of $\mach(m)$ from $m^{1/3}$ for $\mach \rightarrow 1$.

It should be pointed out that the result that it is the
highest--Mach-number shocks that process most energy refers to the
energy processed by accretion shocks alone (which are the shocks
treated in this work). Merger shocks are generally expected to
dominate the overall energetics of cosmic shocks (e.g. Miniati \etal 2000;
Gabici \& Blasi 2003; Ryu \etal 2003), because the density of the
(already virialized) gas
they process is much higher (by a factor between $\sim 20$ and  $200$
on average, depending on the environment in which accretion shocks
lie) than the density of the gas processed by accretion
shocks. However, because the virialized gas relevant to merger shocks 
has also been preheated to
much higher temperatures compared to the gas accreted by accretion
shocks, the Mach numbers associated with merger shocks are
significantly lower. Hence, accretion shocks are expected to dominate
the {\em strong shock} energetics.

The effect of taking into account the environmental density
fluctuations in the primordial density field 
using the double  distribution of collapsed structures is demonstrated
by the curves corresponding to model 2, and is two-fold.
On the one hand, every mass bin is spread out to a larger $\mach$
range, as each accretor mass now corresponds to a distribution of Mach
numbers, reflecting the associated distribution of environmental overdensities.
On the other hand, the local density of the material immediately outside
the accretor is now correlated with the accretor mass. Higher-mass
objects, which have a dominant contribution to the kinetic power and, 
to a lesser extent, the mass processed by accretion shocks,
 tend to reside in higher-mass environments. As a result, the overall
 amplitude of both the kinetic power and the mass current
 distribution is increased.

A most striking environmental effect in the case of the kinetic power
distribution is that, as
redshift decreases, a  second peak separates out 
in model 2, which is absent in the base model. 
The presence of this second, high-$\mach$ 
peak is due to the gradual shift of 
increasingly massive structures towards underdense environments. 
The lower-$\mach$ peak is, conversely, due to the higher-mass structures
which are embedded inside higher-overdensity and hence
higher-temperature environments.

The additional effect of the filamentary structure of the universe is
demonstrated by the curves corresponding to model 3. The inclusion of
filaments results to a shifting of both the mass current and the
kinetic power distributions towards lower Mach numbers (due to the
preheating of the pre-shock gas by filament shocks) and higher
amplitude (due to the additional density enhancement of pre-shock
material inside filaments). At low redshifts, the amplitude of
both distributions shifts by an order of magnitude, reflecting our 
assumed mean filament overdensity. Although our recipe for the
inclusion of the effect of filaments in model 3 is only a first
approximation to the analytical treatment of the issue,
there are two robust conclusions we already draw from this
analysis. First, the effect of filaments is {\em at least} of the
magnitude derived here, since all of our assumptions were chosen so
that they err on the side of underestimating both the
filament-associated preheating as well as density enhancement. Hence,
our results emphasize the need for a detailed model for
filament-associated processes and their cosmic evolution, which have
been shown to be an important factor in determining the statistical
properties of cosmic accretion shocks. Second, the 
{\em direction} of the changes in the mass current and kinetic
power distributions due to filaments will be towards lower Mach
numbers and higher amplitudes, if anything at a higher degree than
predicted here. Even with more sophisticated models, this result is
not expected to change qualitatively. 

Note that the mean filament temperature we have used ($T \sim 10^5 {\,
\rm K}$) is rather on the low-end of the temperature distribution of
filaments in the present-day universe as found in numerical
simulations. Since we do not account for redshift evolution of the
mean filament temperature, we have chosen this low value to ensure
that we do not overestimate the effect of filament preheating on
accretion shocks at higher redshifts. A more realistic mean filament
temperature of $T\sim 10^6 {\rm \, K}$ at low redshifts (right panels
of Figs. \ref{fig:mcdists} and \ref{fig:kindists}) would roughly
result to a horizontal translation of the model 3 curve towards lower
Mach numbers by a factor of $\sim 3$. 

\subsection{Integrated quantities and evolution with cosmic epoch}

\begin{figure*}
\plottwo{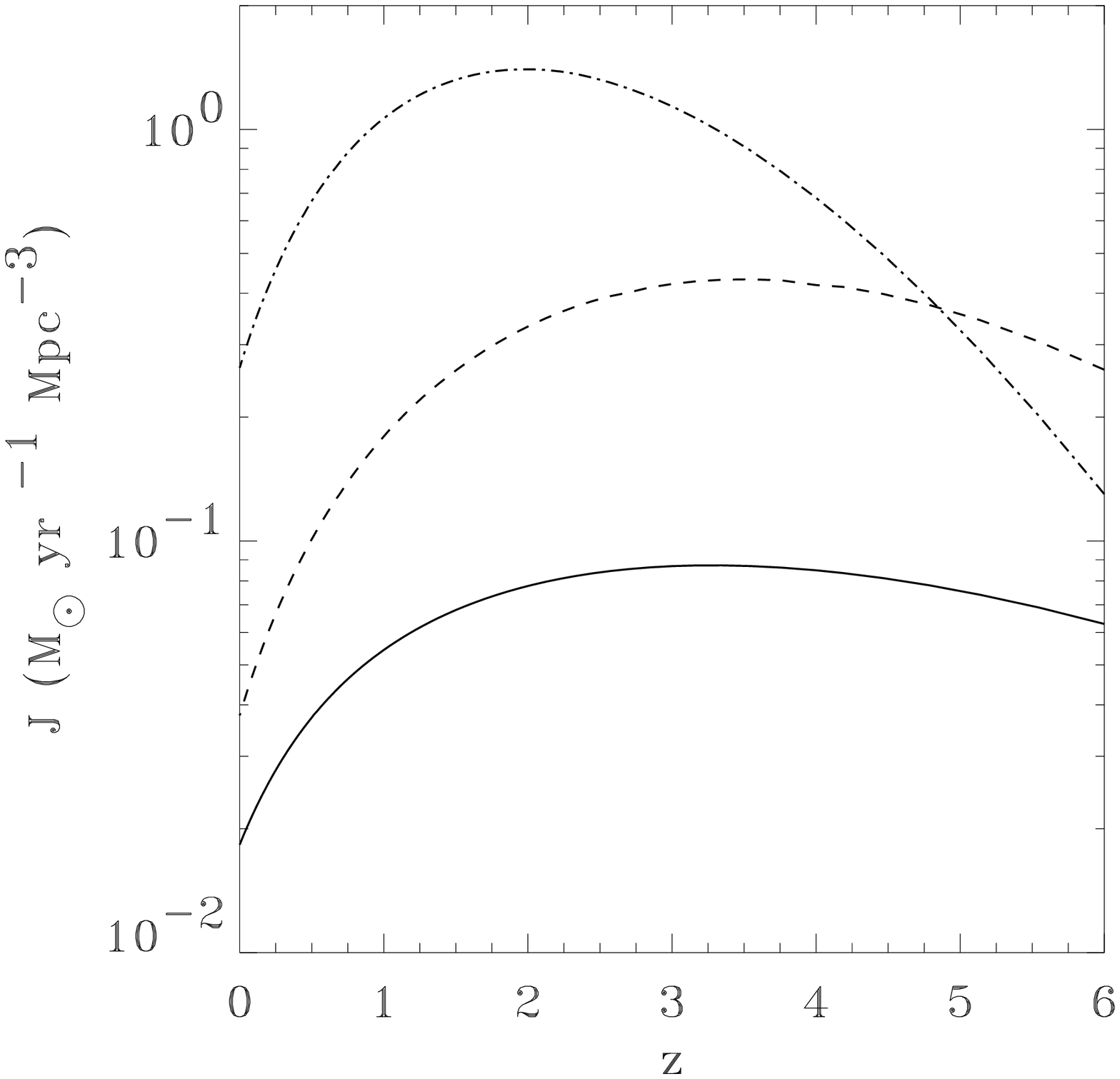}{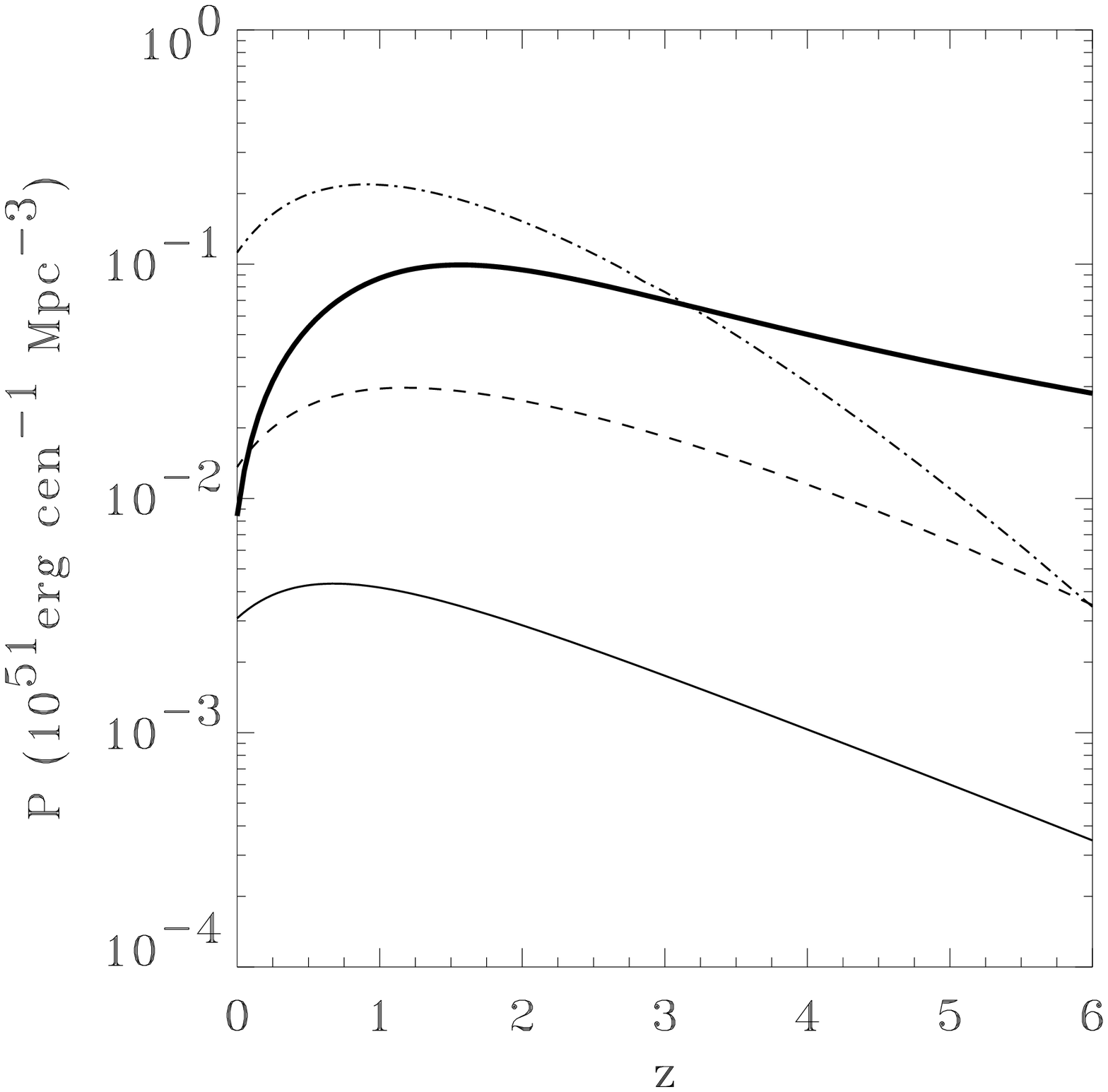}
\caption{\label{fig:JP} Integrated mass current $J$ (left panel) and
  integrated kinetic power $P$ (right panel) as a function of
  redshift. Solid line: model 1 (no environmental effects); 
dashed line: model 2
(effect of primordial density fluctuations); dot-dashed line: model 3
(effects of primordial density fluctuations and filaments).The thick
  solid line in the right panel represents the energy output of
  supernovae as derived from the Cole et al. (2001) fit for the cosmic
  star formation rate.}
\end{figure*}

Figure \ref{fig:JP} shows the evolution of the
integrated mass current $J$ (left panel) and integrated kinetic power
$P$ (right panel) for shocks of any Mach number, and for
redshifts between $6$ and $0$. The solid line corresponds to model 1,
the dashed line to model 2, and the dot-dashed line to the ``full''
model 3. The thick solid line corresponds to the supernova energy
input to the intergalactic medium (which is discussed in detail in the
next section).

The integrated mass current peaks at $z \sim 2$ in the case of model
3, while the models 1 and 2 mass current histories peak closer to 
 $z \sim 3$. The reason for this difference is that
the filament preheating and the associated decrease
of accretion shock Mach number for given accretor mass result to a
certain loss of power from the mass current distribution, as the
lowest mass objects can no longer satisfy $\mach > 1$. The effect is
more pronounced at higher redshifts, when the largest collapsed
objects are less massive than the largest objects in the local
universe. As a result, the evolution of the mass current is more
severe at high redshifts for model 3, and the integrated mass current
peaks at a lower redshift than in the other two models. 

The same effect is present but less pronounced in the case of the
integrated kinetic power: the high-z tail of model 3 falls more
sharply with increasing redshift than the tail of models 1 and 2,
however the effect is not strong enough to affect the location of the
peak, which is found at $z \sim 1$ for all three models. 

The difference in the location of the peak in redshift between
integrated mass current and kinetic power is due to the difference in
the mass dependence of the associated distributions. The kinetic power
distribution depends on mass much more strongly than the mass current,
and thus the integrated kinetic power history ``prefers'' lower
redshifts, where the largest collapsed structures (which process most of
the kinetic power crossing accretion shocks) are more massive. 
For the same reason, the ``loss of power'' due to the presence of
filaments is much milder in the kinetic power history: the
lower-mass structures, which are in danger of ``dropping out'' from
the population of structures hosting shocks in a higher-temperature
environment, do not contribute much to the kinetic power distribution
to begin with.

At redshifts close to their respective peaks, the integrated mass
current and kinetic power as calculated by model 2 are enhanced 
by about an order of magnitude
with respect to the no-environmental-effects model 1. Similarly, the
predictions of model 3 (filaments + primordial overdensities) 
are about an order of magnitude enhanced with respect to those of
model 2 (primordial overdensities only).

\subsection{Comparison of accretion shocks with other intergalactic
  medium energy inputs}

In this section, we will discuss how the kinetic energy processed by
cosmic accretion shocks compares with other energy inputs and
characteristic energy scales of the intergalactic medium. 

In the right panel of fig. \ref{fig:JP} we have plotted, along with
the integrated kinetic power of cosmic accretion shocks as predicted
by models 1, 2 and 3, the energy input of type II supernovae as a
function of redshift (thick solid line). 
The supernova energy history was derived from
the Cole at al (2001) fit to their dust-corrected measurements of the 
cosmic star formation rate, assuming a Salpeter mass function with a
supernova progenitor mass cutoff of $8 {\rm M_\odot}$, and adopting a 
supernova explosion energy of $10^{51} {\, \rm erg}$. The
accretion-shock--processed energy is higher than the energy output of
supernovae for all redshifts $\lesssim 3$. In addition, at low
redshifts the star formation rate decreases with decreasing redshift
much more strongly than the accretion shock processed power, and as a
result in the local universe the energy input to the intergalactic
medium due to accretion shocks
is expected to be (at least) one order of magnitude higher than the
energy output of supernovae. 

>From the point of view of particle acceleration, despite the fact that
the efficiency with which accretion shocks may accelerate high-energy
particles is largely unconstrained, the energy available for particle
acceleration is much larger than in the case of supernovae, and in
this respect the potential of cosmic accretion shocks as sites of
cosmic ray acceleration is once again seen to be promising.

\begin{figure}
\plotone{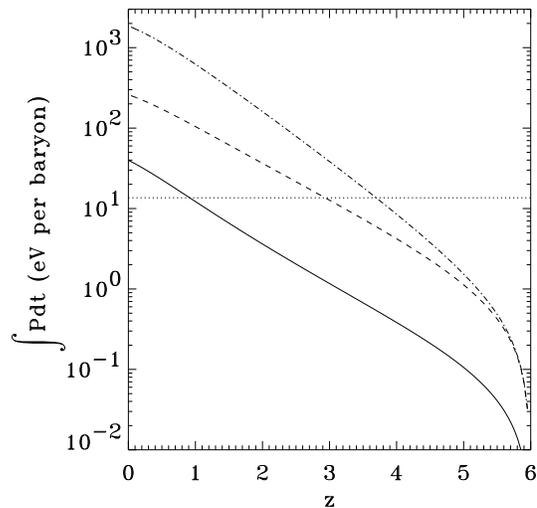}
\caption{\label{fig:CP} Cumulative kinetic power $\int P dt$ (in units
  of eV per baryon in the universe), as a
  function of redshift. Solid line: model 1; dashed line: model 2; 
dot-dashed line: model 3. The horizontal dotted line 
corresponds to 13.6 eV per baryon.}
\end{figure}

Finally, in Fig. \ref{fig:CP} we plot the 
cosmic history of $\int P dt$ (the cumulative processed
kinetic energy) in units of eV per baryon {\em in the universe} (as
opposed to per shocked baryon). Again, the solid line corresponds to 
model 1, the dashed line to model 2 and the dot-dashed line to the
``full'' model 3. 
The horizontal line in this plot corresponds to 13.6 eV per baryon.
>From the location of the intersection of the horizontal line with the
$\int P dt$ curve, we can conclude that by redshift $z \sim 3.5$ (for
model 3), the energy processed by accretion shocks 
{\em alone} would have become comparable to the energy needed to 
reionize the universe\footnote{This is an
  order-of-magnitude estimate, meant to give a feeling of the amount
  of energy processed by shocks as compared to other energy inputs in
  the IGM. If one wanted to consider shocks as an actual reionization
  mechanism, a detailed modeling of the reionization process would be
  required, see e.g. Miniati \etal 2004.}. 

\section{Comparison to past work} 
\label{pastwork}

\subsection{Comparison to baryon energetics from cosmological simulations}

If we integrate $J$ (the mass current through shocks of any Mach
number) as predicted by model 3 (dot-dashed line in left panel of
Fig. \ref{fig:JP}) over time, we find that 
by the present cosmic epoch, $51\%$ of the baryons in the universe
have been processed by accretion shocks, and are therefore residing
within structures of virial temperature roughly $> 10^4 {\rm \, K}$
(mass $\gtrsim 10^{10} {\rm \, M_{\odot}}$). 
We can compare this value to the findings of cosmological
simulations. If we assume that 
$~$1/2 of the baryons of temperatures between $10^4-10^7 {\rm \, K}$
belong to collapsed objects of corresponding virial temperatures, and
we add these baryons to the ``condensed'' (cooled and belonging to
galaxies) and ``hot'' ($T > 10^7 {\rm K}$ and belonging to high-mass
collapsed objects) baryons, we get a total baryon fraction which is
between $35-60\%$, depending on the specifics of each simulation (see,
e.g. simulations in Dav\'{e} \etal 2001). 

As far as the cumulative energy processed by accretion shocks $\int P dt$ is
concerned, the prediction of model 3 for by $z=0$ is 
$P \approx 2 {\rm \, keV}$ per baryon in
the universe (dot-dashed line in Fig. \ref{fig:CP}). 
Note that this is {\em not} a prediction for the mean
per baryon in the universe energy today, since our calculation does
not account for any cooling losses or any other energy inputs to the
IGM. Still, it is interesting to note that it is within a factor of
$\sim 2$ of the value found in cosmological simulations by e.g. Cen \&
Ostriker (1999), who (accounting for cooling as well as star
formation feedback) calculate a density-weighted average temperature
(the quantity representing the mean per-baryon-in-the-universe
energy) of $\sim 10^7 {\rm \, K}$ at $z=0$.

It is therefore very encouraging to see that, given the simplifying
assumptions and idealizations of our model (spherically symmetric
infall, Press-Schechter mass function, absence of time evolution in
filament temperature and density), our results are in reasonable
agreement with the findings of cosmological simulations, indicating
that our most sophisticated model (model 3) has captured the essential 
elements of the gravity-driven, diffuse-matter-accretion--associated 
energetics of cosmic baryons. More importantly, we see that our
other two models, which differ 1-2 orders of magnitude in
their predictions for the cumulative quantities discussed above, are
{\em not} sufficient to reproduce the results of cosmological
simulations. This fact is suggestive of the importance of filaments
not only in processing baryons in their own filament shocks and
giving rise to the warm-hot intergalactic medium, but also
in modifying the energetics of accretion shocks. 

\subsection{Comparison to simulations of cosmic shocks}

\begin{figure*}
\plottwo{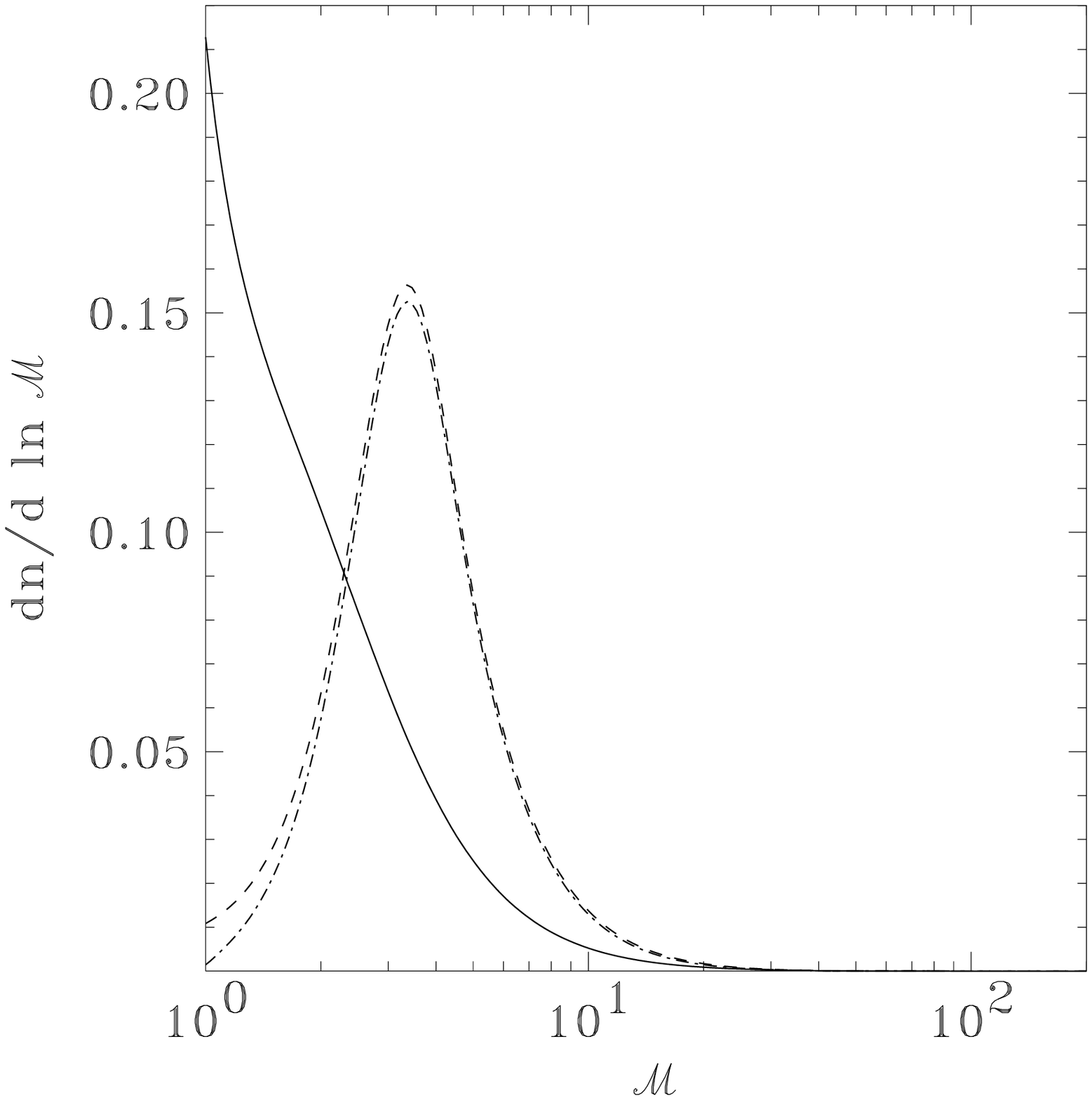}{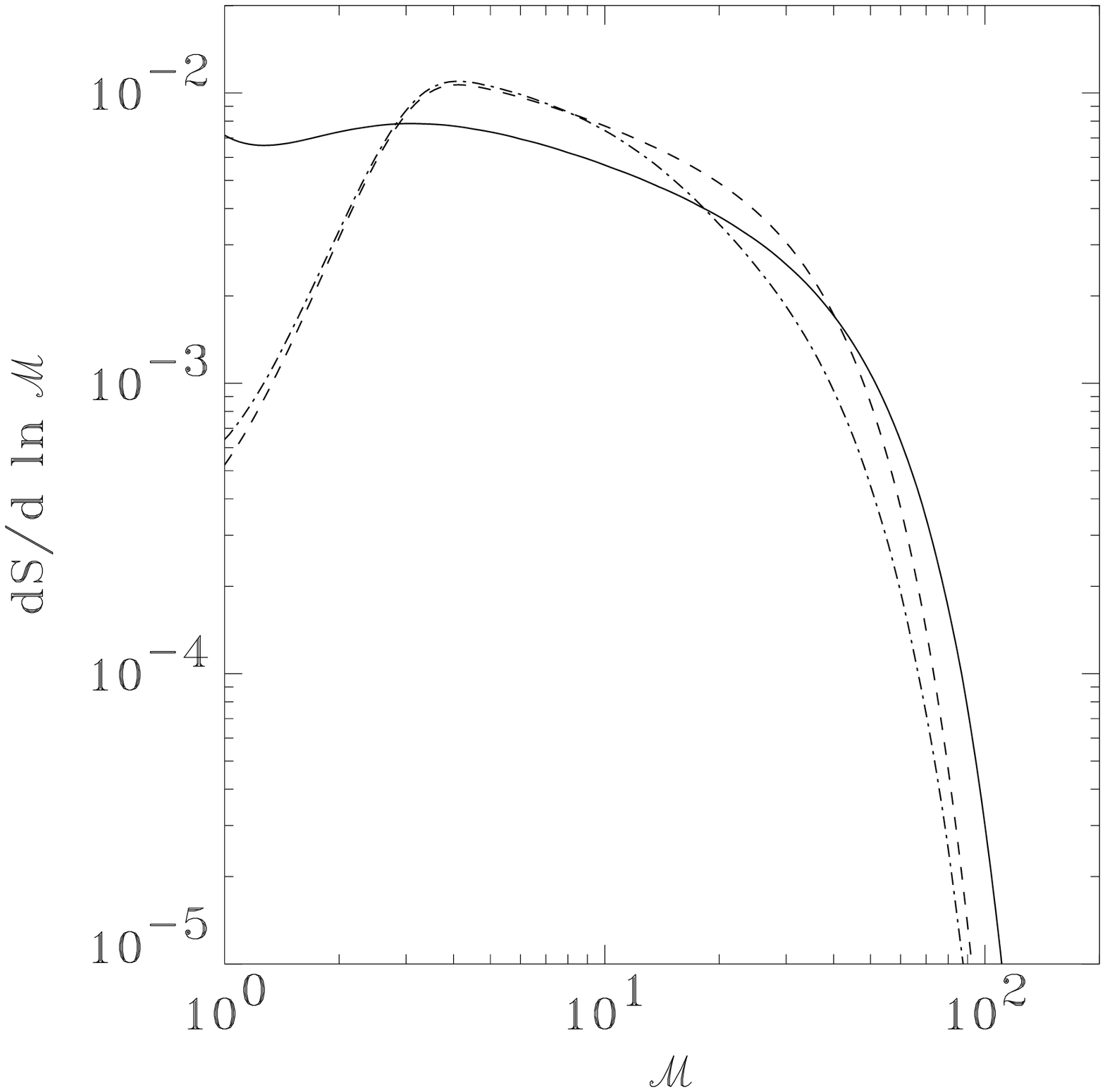}
\caption{\label{fig:MS} Left panel: distribution of the comoving number 
density of accreting structures per
logarithmic Mach number interval of their respective accretion
shocks, $dn/d\ln \mach$, 
with units number of structures per comoving ${\rm Mpc^3}$.
Right panel:
distribution of the comoving shock surface area per
  logarithmic Mach number interval per volume under consideration, 
$dS/d\ln \mach$
with units of ${\rm Mpc}^{-1}$. 
 Solid
line: model 1; dashed line: model 2; dot-dashed line: model 3.
All curves correspond to $z=0$. 
Only structures with 
masses $> 8 \times 10^9 {\rm \, M_\odot}$ are plotted. }
\end{figure*}

Figure \ref{fig:MS} shows two quantities commonly used in the cosmic shock
literature to describe the properties of populations of different
types of cosmic shocks. In the left panel, we plot
the distribution of the comoving number 
density of accreting structures per
logarithmic Mach number interval of their respective accretion
shocks, $dn/d\ln \mach$, 
with units number of structures per comoving ${\rm Mpc^3}$, for the
three models we have presented in this work and for $z=0$. 
In the right panel, we
plot the distribution of the comoving shock surface area per
  logarithmic Mach number interval per volume under consideration, 
$dS/d\ln \mach$
with units of ${\rm Mpc}^{-1}$ (ratio of shock surface over space
volume). Only structures with virial temperatures $>10^4 {\rm \, K}$
(masses $> 8 \times 10^9 {\rm \, M_\odot}$) are plotted. 
As expected, these distributions are dominated by the low-mass,
low--Mach-number structures (the same analytic scalings used in \S
\ref{analarg} give, for the low-mass behavor of 
our no-environmental-effects model 1 
$dn/d\ln \mach \propto m^{-1}$ and $dS/d\ln \mach
\propto m^{-1/3}$). 
The number distribution in the Press-Schechter--based model (model 1) 
monotonically increases for decreasing $\mach$ 
since in this case there is a one-to-one correspondence between mass and Mach
number.  Hence the number distribution of objects simply follows the
Press-Schechetr mass function modulated by the mass-Mach conversion.
In the case of the double-distribution--based models (models 2 and 3), 
the mass cutoff, in combination with the
distribution of pre-shock sound speeds for each
accretor mass described by the double distribution, results in a 
number distribution which peaks at
$\mach \lesssim 4$, a position which is defined by the combination of
the virial temperature at the mass cutoff (which is the most populated
available mass bin) and the sound speed
corresponding to the most probable environment at that particular mass. 
Note that the difference in both the number and surface distributions
between models 2 and 3 (without and with the effect of filaments) 
is very small. This is because we have used the double distribution of
cosmic structures to predict which structures are located inside
filaments, and the double distribution predicts that most of the
low-mass cosmic structures (the ones dominating the number and surface
distributions) are found inside underdense regions, and hence not inside
filaments. 

For the case of {\em accretion shocks} which we
have studied here, we can then draw the following conclusions:
\begin{enumerate}
\item The energetics of these shocks are determined by the few,
      high--Mach-number, high-mass structures, and are not
      significantly affected by the majority of structures, which have
      low masses and low accretion shock Mach numbers. 
\item Our results are in natural agreement with the findings of 
Ryu \etal (2003), who performed a series of simulations with increasing mass
and spatial resolution, and found convergence for their
results on kinetic power distribution (dominated by high-mass
structures) but not for the surface
distribution (dominated by low-mass structures). 
However, since the physical impact of shocks
to their environment is better represented by the kinetic power 
distribution, which is not sensitive to the behavior of
low-mass structures, the lack of convergence in the
number-dominated distributions in cosmological simulations 
should not decrease confidence in the relevance and robustness 
of the conclusions of such cosmological simulations relating to the
effect of accretion shock on the intergalactic medium.
\end{enumerate}

Finally, we should caution the reader that a direct comparison
between our Fig. (\ref{fig:MS}) and similar plots in cosmic shock 
literature is not
straight-forward - the shocks treated here are {\em
exclusively} accretion shocks, and they do not include merger or
filament shocks. When shock surfaces in the universe are labeled 
as internal or external, accretion shocks are split between the two
types: internal shocks can be merger shocks, or they can be accretion shocks
associated with structures found inside filaments; external shocks on
the other hand can be accretion shocks associated with structures
which are not located inside filaments, or they can be filament
shocks. 

\subsection{ Comparison to other analytic models of cosmic 
shock populations}

Press-Schechter extensions have been used extensively to model different
``families'' of large-scale cosmic shocks, with emphasis primarily on
the emission properties of such populations, rather than the
properties of the shocks themselves. 

Waxman \& Loeb (2000) assumed a Press-Schechter mass function to 
describe an underlying population of accretors hosting particle
accelerating shocks which would produce, through their radio and
gamma emission, fluctuations in the radio and gamma-ray
background. All shock quantities required for their
emissivity and fluctuation calculations (mass accretion rate, shock
temperature, shock radius) were derived as functions of
the accretor velocity dispersion up to dimensionless constants, left to
be calibrated against simulations. Of all the analytic shock models
discussed in this section, the Waxman \& Loeb (2000) model is the one
with the most similarities with the ``base'' model we use in this work (the
model containing no environmental effects). 
The two primary similarities are that
a) we also take the 
underlying accretor distribution to be described by a 
Press-Schechter mass function and b) the mass-scaling of the 
shock properties is the same over a large range of masses. 
In our case, the shock properties are obtained the temperature 
jump across the shock, which involves
the mass-dependent virial temperature of each object and some
environmental temperature. At the 
large-temperature-jump/high-Mach-number limit, 
the mass dependencies of shock properties in our model
agree with those of Waxman \& Loeb (2000), as the
velocity dispersion and the free-fall velocity around an object of a 
certain mass (which is encoded in the accretion shock Mach number
provided that Mach $\gg 1$, or, equivalently, that the gas pressure
effects in the pre-shock gas are negligible) 
necessarily have the same mass dependence. 

However, our ``base model'' describes different aspects of the
accretion shock population that Waxman \& Loeb (2000), as
they calculate the shock
gamma and radio emission properties while we calculate the mass 
current and shock energetics, exhibit their dependence on 
Mach number, calculate the cosmic distribution of mass and energy flows
as a function of Mach number, and explicitly construct 
histories for the accreted mass and energy.

Fujita \& Sarazin (2001) used a version of the Lacey \& Cole (1993)
merger formalism (a Press-Schechter extension) to describe mergers and
accretion in a unified way (in their picture, accretion is the sum of
all minor mergers for which the relative mass increment of the
accreting halo does not exceed some threshold). They as well use their
model to obtain the nonthermal emission properties of
merging and accreting clusters, and they compare how such properties
differ between clusters which have recently undergone some major
merger and clusters which have been only accreting through minor
mergers. Treating accretion as a subcase of the general merger picture
associated with hierarchical structure formation is a widely used
technique. It was adopted also by  Gabici \& Blasi (2003a, b) who also
used a semi-analytic model based on the Lacey \& Cole (1993) merger
formalism to calculate the Mach numbers of shocks resulting from
mergers, the particle acceleration properties of such shocks, and the
contribution of merger-shock--accelerated particles to nonthermal
radiation from large-scale structure (including cluster radio halos
and a contribution to the gamma-ray background)\footnote{Note however 
that Gabici \& Blasi (2004) did make an explicit distinction between 
mergers and diffuse gas accretion when assessing the detectability of
gamma rays from galaxy clusters; in this work, they considered all
accretion shocks to have $\mach \gg 1$, while the scaling of
shock energy with accretor mass was similar to that of 
Waxman \& Loeb (2000).}. 

Such a mergers-only approach is fundamentally rather than
just formally different than the one we use here. First of all,
since in the mergers-only picture all matter is locked up
in collapsed objects of various masses, the
pre-shock material has a mixture of temperatures, characteristic of 
the virial temperatures (and hence masses) of the accumulated halos,
rather than of the diffuse environment of the object. In addition, the
mass accretion rate for a halo of mass $m$ 
in the mergers-only picture is
not $\propto m$ as in the simple, gravitationally-driven accretion of
diffuse gas picture we adopt, but is dependent on the index $n$ of the
power spectrum of initial perturbations (and is $\propto m^{7/6}$ for
$n = 1$). This difference stems from the different physical processes
followed in each case. In the mergers-only, Press-Schechter--extension
picture, growth of structure, whether through accretion or through
mergers, occurs due to the gradual turnaround and collapse of
increasingly large scales. This process is caused by the gravitational
enhancement of perturbations {\em already imprinted} in the
primordial density field. 
The gravitational effect of a
collapsed object within such a larger collapsing structure is simply
to contribute to the mean density of the collapsing region. In our
picture, accretion is driven {\em locally}, 
through the gravitational attraction that
a collapsed structure exerts on its surrounding diffuse gas. 
Each collapsed structure would accrete gas even if it lived inside an
otherwise homogeneous universe.
In nature, both processes contribute to the growth 
of any particular
collapsed structure, and it is hence important to construct 
analytic models
based on both pictures, and gauge the relevant importance of each
process through comparison with observations and simulations. 

More recently, Inoue \& Nagashima (2005) used the  Somerville \&
Kolatt (1999) multiple merger model to follow the population of cosmic
shocks, calculate the associated gamma-ray emission and 
investigate the possibility to use gamma-ray observations as a tool
for the study of the warm-hot intergalactic medium. 
The Somerville \& Kolatt (1999) merger model is an improved version of
a Press-Schechter--based semi-analytic merger-tree construction
algorithm. Its major advantage with respect to other algorithms for
construction of merger trees of present-day objects is that it
predicts a distribution of mass progenitors which is consistent with
the extended Press-Schechter predictions while it enforces strict mass
conservation. This model accounts for some amount of diffuse accreted
matter, which is however 
calculated not by gravitational arguments,
but again as that amount of matter corresponding to progenitors 
less massive
than some pre-set cutoff. 

Finally, Furnaletto \& Loeb (2004) used a modified
Press-Schechter formalism to describe large-scale
shocks that may appear when overdense perturbations reach 
and exceed their turnaround point. The shocks they consider are
located at the boundaries between regions which have reached their
turnaround point, and the background (still expanding) universe.
The shock velocity is derived from a scaling argument to be
approximately equal to the size the shocked region would have had if
it had expanded with the Hubble flow, divided by the age of the
universe. All shocks in this work are assumed to be strong, and 
Mach-number dependencies are not investigated. 
>From a physical standpoint, the shocks described by these
authors occur much earlier in the evolution of each structure than the
accretion shocks we describe here, and they are the shocks responsible
for heating up the bulk baryonic component of collapsed objects from
mean intergalactic medium temperatures to the virial temperature of
structures. In our shock classification scheme, the Furlanetto \& Loeb
(2004) shocks would be the gravitationally-driven component of
filament shocks. 

To summarize, our model of accretion shocks is different from
and complementary to all of the exiting treatments.  Our
focus on the role of diffuse gas accretion is qualitatively
and quantitatively different from the mergers-only approaches,
and these differences can be exploited to test against
the physical reality in which both accretion modes occur.
Our model is most similar to that of Waxman \& Loeb (2000), but
we extend the Press-Schechter treatment to include both
the effects of environment, and to address the effects
of preheating by filament shocks.  Finally, our work differs
from much of the existing literature by focusing on the
shock energetics, and to our knowledge is the first to
explicitly calculate the net cosmic processing of both
energy and mass through accretion shocks over cosmic
history.

\section{Discussion}\label{disc}

In this paper, we have investigated analytically
the properties of cosmic accretion shocks around collapsed structures.
We have calculated the mass and kinetic energy currents
as a function of Mach number for accretion shocks on
individual structures of arbitrary mass at any epoch.
Using this, we have computed the
cosmic distribution of mass and energy flows as a function
of Mach number, and we have
calculated the evolution of these quantities as a function
of redshift.

We have found that environmental factors play a major role
in shaping the qualitative and quantitative properties of
accretion shocks.  We demonstrated the impact of environmental
effects by exploring three models for
the cosmic shock population. The first used the Press-Schechter mass
function to describe the underlying population of collapsed, accreting
objects. All such objects were assumed to accrete material of the same
density and temperature. This was our ``control'' model, which did not
include any environmental effects. The second model used the
double distribution of collapsed structures (Pavlidou \& Fields 2005)
to describe the
distribution of accreting objects with respect to both their mass and
local environment overdensity or underdensity. The overall mass
distribution of objects is the same as in the first model, as the
double distribution integrates to the Press-Schechter mass function. 
This model then only included environmental effects caused by
fluctuations in the primordial density field. Our third model was a
refinement of model 2, which considered all objects predicted by the
double distribution to reside inside overdensities to accrete gas
preheated and compressed in cosmic filaments. 

The inclusion of primordial density fluctuations resulted to broader
distributions with respect to Mach number of the mass current and
kinetic power processed by accretion shocks, and an overall increase
in the amplitude of such distributions, owing to an increase in the
density of accreted material by the most massive objects, which
dominate especially the processed kinetic power. 
Perhaps the most striking effect of accounting for this environmental
factor is the separation of a second, high Mach number peak in the
kinetic power distribution at low redshifts, due to an increasing
number of higher-mass structures concentrating inside
underdensities. 

The inclusion of filament preheating and compression further increased
the amplitude of the mass current and kinetic power distributions, and
shifted the distributions towards lower Mach numbers. A more refined
filament model, the need of which is underlined by our results, 
is expected to enhance these trends even more.
Our treatment of both environmental factors
has been systematically conservative in
estimating the impact of the effects on the shock properties.
Consequently, we expect that these differences
will be evident at least qualitatively in full numerical models.

The integrated kinetic power processed by shocks peaks at a redshift of
$\sim 1$, while 
the integrated mass current peaks at higher
redshifts, $z\sim 2$ (for model 3).
The effect of the local environment is to increase the
overall level of the processed energy at peak redshifts by a factor of
$\sim 10$ for each environmental factor included.

Comparing the energy processed by cosmic accretion shocks to other
energy inputs to the intergalactic medium, we find that the energy
input of accretion shocks (as predicted by model 3) 
is higher that that of Type II supernovae
for all $z \lesssim 3$, and it becomes more than an order of magnitude
higher in the local universe. In addition, we 
found that energy processed by accretion shocks alone 
becomes comparable to the energy needed to reionize the universe by 
$z \sim 3.5$.

Numerical and observational tests of our model will quantify the
importance of accretion shocks and their prominence among cosmic shock
mechanisms.  Our emphasis on accretion due to diffuse matter is
complementary to minor mergers, thus comparison with simulations
will shed light on the relative importance of these two components in
the $\Lambda$CDM scenario.  To compare the present results with
simulation is however neither trivial nor immediate, because
simulations include merger and filament shocks as well, but
moreover because accretion shocks as we have defined them do not
uniquely map onto the classification schemes used in the existing
literature. 
A detailed, self-consistent comparison with simulations will appear in
future work.

Of course, the question of the nature of cosmic shocks and the
importance of accretion processes ultimately will be resolved observationally.
Galaxy clusters provide a promising site to study accretion shocks
(e.g. Lieu \etal 1996; Ensslin \etal 1998; Fusco-Femiano \etal 1999; 
Fujita \& Sarazin 2001; Govoni \etal 2001; Miniati \etal 2001a,b; 
Bagchi \etal 2002; Berrington \& Dermer 2003; 
Feretti \etal 2004; Gabici \& Blasi 2004; 
Keshet \etal 2004; Kocsis \etal 2005)
As with simulations, there is a need to separate merger
versus accretion components;
this might be done spatially (e.g. Inoue \etal 2005), with
accretion shocks dominating
peripheral emission, particularly that due to inverse Compton. 
In addition, detection of
 $\gamma-$ray emission unambiguously associated with structure
 formation shocks, either by TeV (such as CANGAROO, HESS, MAGIC  
and VERITAS) or GeV
 (such as AGILE and GLAST) gamma-ray telescopes, will provide 
invaluable insight
not only in the nature of the accretion process itself, 
but also the currently elusive subjects of the acceleration efficiency 
associated with cosmic shocks, as well as the properties of
large-scale magnetic fields.

\acknowledgments{We thank Tom Abel, Susumu Inoue, Tom Jones, 
Hyesung Kang, Francesco Miniati, Temelachos Mouschovias, 
Tijana Prodanovi\'{c}, Kostas Tassis, Ben Wandelt,
Ellen Zweibel, and an anonymous referee for useful discussions. 
This work was supported by the
National Science Foundation through grant AST-0092939.
The work of V.P. was partially supported by Zonta International through 
an Amelia Earhart Fellowship. V.P. additionally acknowledges support
by the Kavli Institute for Cosmological Physics through 
grant NSF PHY-0114422.}


\begin{thebibliography}{}

\bibitem[Bagchi et al.(2002)]{2002NewA....7..249B} Bagchi, J., En{\ss}lin, 
T.~A., Miniati, F., Stalin, C.~S., Singh, M., Raychaudhury, S., \& 
Humeshkar, N.~B.\ 2002, New Astronomy, 7, 249 

\bibitem{BerrD03}
Berrington, R.~C. \& Dermer, C.~D. 2003, \apj, 594, 709

\bibitem{bert85}
Bertschinger, E. 1985a, \apjs, 58, 39

\bibitem{bertV}
Bertschinger, E. 1985b, \apjs, 58, 1

\bibitem{BSFG01}
Brunetti, G., Setti, G., Feretti, L. \& Giovannini, G. 2001, NewA, 6, 1

\bibitem{BBCG04}
  Brunetti, G., Blasi, P., Cassano, R. \& Gabici,S. 2004, \mnras, 350, 1174

\bibitem{CO99}
Cen, R. \& Ostriker, J.~P. 1999, \apj, 514, 1

\bibitem{Cyb03}
Cyburt, R.~H., Fields,B.~D. \& Olive, K.~A. 2003, PhLB 567, 227

\bibitem{D01}
Dav\'{e} R. {\it et al.} 2001, \apj, 552, 473

\bibitem[Ensslin et al.(1998)]{1998A&A...332..395E} Ensslin, T.~A., 
Biermann, P.~L., Klein, U., \& Kohle, S.\ 1998, \aap, 332, 395 

\bibitem{FBC02}
Fang, T., Bryan, G.~L., Bryan, \& Canizares, C.R. 2002, \apj, 564, 604

\bibitem[Feretti et al.(2004)]{2004NewAR..48.1137F} Feretti, L., Burigana, 
C., \& En{\ss}lin, T.~A.\ 2004, New Astronomy Review, 48, 1137 

\bibitem{finog03}
Finoguenov, A., Briel, U.~G. \& Henry, J.~P. 2003, \aap, 410, 777

\bibitem{fs01}Fujita, Y., \& Sarazin, C.~L.\ 2001, \apj, 563, 660 

\bibitem{Fuk98}
Fukugita, M., Hogan, C.~J. \& Peebles, P.~J.~E. 1998, \apj, 503, 518

\bibitem{furlanetto}
Furlanetto, S.~R. \&  ~Loeb, A. 2004,\apj 611, 642

\bibitem{GB03p}
Gabici, S. \& Blasi, P. 2003a, \apj, 583, 695

\bibitem{GB03g}
Gabici, S. \& Blasi,P. 2003b, APh, 19, 679

\bibitem{GB04}
Gabici, S. \& Blasi, P. 2004, APh, 20, 579

\bibitem[Govoni et al.(2001)]{2001A&A...376..803G} Govoni, F., Feretti, L., 
Giovannini, G., B{\"o}hringer, H., Reiprich, T.~H., \& Murgia, M.\ 2001, 
\aap, 376, 803 

\bibitem{HO86}
Hegyi, D.~J. \& Olive, K.~A. 1986, \apj, 303, 56

\bibitem{Hel98}
Hellsten, U., Gnedin, N.~Y. \& Miralda-Escud\'{e}, J. 1998, \apj 
509, 56

\bibitem{inoue05} 
Inoue, S. \& Nagashima, M. 2005, AIP Conf.~Proc.~745: 
High Energy Gamma-Ray Astronomy, 745, 567 

\bibitem {ias05} 
Inoue, S., Aharonian, F.~A., \& Sugiyama, N.\ 2005, \apjl, 628, L9 
 

\bibitem{KJ05}
Kang, H. \& Jones, T.~W. 2005, \apj, 620, 44

\bibitem{KRCS05}
Kang, H., Ryu, D., Cen, R. \& Song, D. 2005, \apj 620, 21

\bibitem{kesh} 
Keshet, U., Waxman, E., Loeb, A., Springel,V.  \&
Hernquist, L. 2003, \apj, 585, 128

\bibitem[Keshet et al.(2004)]{2004NewAR..48.1119K} Keshet, U., Waxman, E., 
\& Loeb, A.\ 2004, New Astronomy Review, 48, 1119 

\bibitem[Kocsis et al.(2005)]{2005ApJ...623..632K} Kocsis, B., Haiman, Z., 
\& Frei, Z.\ 2005, \apj, 623, 632 

\bibitem{KBH}
Kuo, P.~H., Bowyer, S. \& Hwang, C.-Y. 2005, \apj, 618, 675

\bibitem{lc93} 
Lacey, C. \& Cole, S. 1993, \mnras, 262, 627

\bibitem[Lieu et al.(1996)]{1996ApJ...458L...5L} Lieu, R., Mittaz, 
J.~P.~D., Bowyer, S., Lockman, F.~J., Hwang, C.-Y., \& Schmitt, 
J.~H.~M.~M.\ 1996, \apjl, 458, L5 

\bibitem{lw}
Loeb, A. \& Waxman, E. 2000, Nature, 405, 156

\bibitem{Mat03}
Mathur, S., Weinberg, D.~H. \& Chen, X.2003, \apj, 582, 82

\bibitem{min_shock}
Miniati, F., Ryu, D., Kang, H., Jones, T.~W., Cen, R. \&  
Ostriker, J.~P. 2000, \apj, 542, 608

\bibitem{min01}
Miniati, F., Jones, T.~W., Kang, H. \& Ryu,D. 2001a,  \apj 562, 233

\bibitem{min01b}
Miniati, F., Ryu, D., Kang, H. \& Jones, T.~W. 2001b, \apj, 559, 59

\bibitem{min02} 
Miniati, F. 2002, \mnras, 337, 199

\bibitem{min03} 
Miniati, F. 2003, \mnras, 342, 1009

\bibitem{min04} 
Miniati, F., Ferrara, A., White, S.~D.~M., \& Bianchi, S.
2004, \mnras, 348, 964 
 

\bibitem{ns01} 
Nath, B.~B., \& Silk, J.\ 2001, \mnras, 327, L5 

\bibitem{Nic02}
Nicastro F. {\it et al.} 2002, \apj, 573, 157

\bibitem{nicastro}
Nicastro F. {\it et al.} 2005, Nature, 433, 495

\bibitem{pfd} 
Pavlidou, V. \& Fields, B.D. 2005, \prd, 71, 043510

\bibitem{PL98}
Perna R. \& Loeb, A. 1998, \apj, 503, L135

\bibitem{ps74} 
Press, W.H. \&  Schechter, P. 1974, \apj, 187, 425

\bibitem{ProdF04}
Prodanovi\'{c}, T. \& Fields, B.~D. 2004, \apj, 616, L115

\bibitem{ProdF04b}
Prodanovi\'{c}, T. \& Fields, B.~D. 2005, \apj, 623, 877

\bibitem{RPSM}
Reimer, O., Pohl, M., Sreekumar P. \& Mattox, J.~R. 2003, \apj, 588,

\bibitem{ryu97}
Ryu, D. \& Kang, H. 1997, \mnras, 284, 416

\bibitem{RKJ03}
Ryu, D., Kang, H., Hallman E. \& Jones, T.~W. 2003, \apj, 593, 599

\bibitem{SchMuk} 
Scharf, C.~A. \& Mukherjee,R. 2002,  \apj, 580, 154

\bibitem{Sper03}
Spergel, D.~N. {\it et al.} 2003, \apjs, 148, 75

\bibitem{SuIn02}
Suzuki, T.~K. \& Inoue, S. 2002, \apj, 573, 168

\bibitem{SuIn04}
Suzuki, T.~K. \& Inoue, S. 2004, PASA, 21, 148

\bibitem {tk} 
Totani, T. \& Kitayama, T. 2000, \apj, 545, 572

\bibitem{ti}
Totani, T. \&  Inoue, S. 2002, APh, 17, 79

\end{thebibliography}
\end{document}